\documentclass[useAMS,usenatbib]{mn2e}

\usepackage[psamsfonts]{amssymb}
\usepackage[dvips]{graphicx}
\usepackage{amsmath,alltt}                                                                                                                          
\usepackage{multirow}
\usepackage{rotating}
\usepackage{lscape}

\title[The Effects of Orbital Inclination on Star Clusters]{The Effects of Orbital Inclination on the Scale Size and Evolution of Tidally Filling Star Clusters}
\author[Webb, J. J.]{Jeremy J. Webb$^1$, Alison Sills$^1$, William E. Harris$^1$, Jarrod R. Hurley$^2$
\thanks{E-mail: webbjj@mcmaster.ca (JW)} \\
$^1$ Department of Physics and Astronomy, McMaster University, Hamilton ON L8S 4M1, Canada \\
$^2$ Centre for Astrophysics and Supercomputing, Swinburne University of Technology, P.O. Box 218, VIC 3122, Australia}

\begin{document}

\pagerange{\pageref{firstpage}--\pageref{lastpage}} \pubyear{2043}

\maketitle

\label{firstpage}

\begin{abstract}
We have performed $N$-body simulations of tidally filling star clusters with a range of orbits in a Milky Way-like potential to study the effects of orbital inclination and eccentricity on their structure and evolution. At small galactocentric distances $R_{gc}$, a non-zero inclination results in increased mass loss rates. Tidal heating and disk shocking, the latter sometimes consisting of two shocking events as the cluster moves towards and away from the disk, help remove stars from the cluster. Clusters with inclined orbits at large $R_{gc}$ have decreased mass loss rates than the non-inclined case, since the strength the disk potential decreases with $R_{gc}$. Clusters with inclined and eccentric orbits experience increased tidal heating due to a constantly changing potential, weaker disk shocks since passages occur at higher $R_{gc}$, and an additional tidal shock at perigalacticon. The effects of orbital inclination decrease with orbital eccentricity, as a highly eccentric cluster spends the majority of its lifetime at a large $R_{gc}$. The limiting radii of clusters with inclined orbits are best represented by the $r_t$ of the cluster when at its maximum height above the disk, where the cluster spends the majority of its lifetime and the rate of change in $r_t$ is a minimum. Conversely, the effective radius is independent of inclination in all cases.

\end{abstract}



\section{Introduction \label{Introduction}}

The gravitational dynamics of a three-body system which consists of a star orbiting in the combined potential of a star cluster and its host galaxy becomes increasingly complicated as one attempts to make the system more realistic. By treating all three members of the system (star, cluster, galaxy) as point masses, one can easily determine the tidal radius $r_t$ (or Jacobi radius $r_J$) of the cluster, which is defined as the distance beyond which a star feels a stronger acceleration towards the host galaxy than the cluster itself \citep{vonhoerner57}. A straightforward derivation of $r_t$ yields a function that depends on cluster mass $M_{cl}$, galaxy mass $M_g$, and the cluster's galactocentric distance $R_{gc}$:

\begin{equation}\label{rtHoerner}
r_t \simeq R_{gc}(\frac{M_{cl}}{2M_g})^{1/3}
\end{equation}

Allowing the host galaxy to have a non-point-mass potential introduces significant complexity that has led to multiple analytic definitions of $r_t$ \citep[e.g.][]{king62, innanen83, jordan05, binney08, bertin08}. However all analytic expressions of $r_t$, no matter how complex the tidal field, are limited by the assumptions that the host galaxy has a spherically symmetric potential and the cluster has a circular orbit. Under these assumptions the tidal field experienced by the cluster can be taken to be static. The derivation by \citet{bertin08} is likely the most generalized derivation of $r_t$, as spherical symmetry is the only assumption it makes. In that work, $r_t$ is defined as:

\begin{equation} \label{rt}
r_t=(\frac{GM_{cl}}{\Omega^2\upsilon})^{1/3}
\end{equation}

\noindent where $\Omega$, $\kappa$ and $\upsilon$ are:

\begin{equation}
\Omega^2=(d\Phi_G(R)/dR)_{R_{gc}}/R_{gc}
\end{equation}
\begin{equation}
\kappa^2=3\Omega^2+(d^2\Phi_G(R)/dR^2)_{R_{gc}}
\end{equation}
\begin{equation}\label{upsilon}
\upsilon=4-\kappa^2/\Omega^2
\end{equation}

\noindent Here $\Phi_G$ is the galactic potential, $M$ and $R_{gc}$ are the mass and galactocentric distance of the cluster respectively, $\Omega$ is its orbital frequency, $\kappa$ is the epicyclic frequency of the cluster at $R_{gc}$, and $\upsilon$ is a positive dimensionless coefficient. 

Since disk and triaxial elliptical galaxies have non-spherically symmetric potentials, and most Galactic globular clusters have non-circular orbits \citep{dinescu99, dinescu07,dinescu13}, assuming that a cluster experiences a static tidal field as it evolves is clearly incorrect. Various works have studied the evolution of clusters in non-static tidal fields \citep{baumgardt03, giersz09, giersz11, renaud11, webb13,brockamp14,madrid14}, where the easiest approach is to first consider clusters with eccentric orbits in a spherically symmetric tidal field. These studies have shown that a cluster on an eccentric orbit will lose mass faster than if it has a circular orbit at apogalacticon $R_a$, but slower than if it has a circular orbit at perigalacticon $R_p$. The increased mass loss rate is attributed to tidal shocks during perigalactic passes and tidal heating.

Tidal heating and tidal shocks occur when a cluster experiences a time varying gravitational force. A tidal shock refers to a highly varying gravitational force experienced over a short period of time (e.g. perigalactic pass or passage through a disk). During a tidal shock, individual stars undergo an increase in energy that is dependent on their location within the cluster, and the cluster's binding energy is reduced. The orbits of stars during a shock can receive a significant kick, which can push loosely bound stars outside $r_t$ \citep{gnedin97}. Tidal heating on the other hand refers to a slowly varying gravitational force experienced over a long period of time (e.g. eccentric orbit or non-spherically symmetric potential). While the amount of energy injected into the cluster per unit time is much smaller than a tidal shock, over significant periods of time tidal heating can also have a strong influence on a cluster's evolution. Both mechanisms can provide stars with additional energy to escape the cluster that otherwise would remain bound, accelerating mass loss \citep{webb13, brockamp14}.

For a cluster on an eccentric orbit in a spherically symmetric potential, Equation \ref{rt} represents the instantaneous $r_t$ of the cluster, which fluctuates between a maximum at $R_a$ and minimum at $R_p$. In \citet{webb13} we demonstrate that the limiting radius $r_L$ of a tidally filling cluster (the radius at which the stellar density approaches zero) traces $r_t$ at all phases of its orbit. The agreement between $r_L$ and $r_t$ can be attributed to the cluster recapturing stars as it moves away from $R_p$, as well as energy injection from tidal shocks and tidal heating energizing bound stars to larger orbits within the cluster.

To better reflect the globular cluster population of disk galaxies, including the Milky Way, orbits need to be considered that are both eccentric and inclined to the plane of the disk. Studies of the effects of an inclined orbit on star clusters have primarily been focused on the effects of disk shocking as the cluster passes through the plane of the disk.

accelerate mass loss \citep{gnedin97, gieles07, donghia10, madrid14}. 

A cluster on an inclined orbit will not only undergo tidal shocks, but tidal heating as well even though its orbit is circular. If the cluster orbit is inclined \textit{and} eccentric, which is the case for Galactic globular clusters, it will experience a third tidal shock at $R_p$. The overall effect on the mass loss rate and scale size of such a cluster has not been fully explored. 

The purpose of this study is to both isolate and identify the effects of orbital inclination on the evolution of a star cluster and consider the combined effects of orbital inclination and eccentricity. Model N-body clusters with a range of orbital inclinations and eccentricities are evolved from t=0 to 12 Gyr in a Milky Way-like potential. In Section 2 we introduce the models and their initial conditions. In Section 3 we focus on how orbital inclination and eccentricity influence the evolution of cluster mass (M), $r_t$, velocity dispersion $\sigma_V$, $r_L$ and half-mass radius $r_m$. We discuss the results of all our N-body models in Section 4. Specifically we suggest a method to correct both the dissolution time and the theoretical calculation of a cluster's scale size for inclination and/or eccentricity. We summarize our conclusions in Section 5.

\section{The Models \label{stwo}}

Model clusters are evolved from t=0 to 12 Gyr with the NBODY6 direct N-body code (Aarseth 2003). The initial mass of each model is $6 \times 10^ 4 M_\odot$ and has 96000 single stars and 4000 binaries. Stellar masses are drawn from a \citet{kroupa93} initial mass function between $0.1 M_\odot$ and $30 M_\odot$, with each star assigned a metallicity of $Z=0.001$. The distribution of stellar positions and velocities follow a Plummer density profile with a cutoff at $10 r_m$ \citep{plummer11,aarseth74}. The initial half mass radius of each model is set to 6 pc, which ensures that all models exhibit bound stars at or beyond $r_t$ and can be considered tidally filling.

To first study the effects of orbital inclination on star clusters, we simulate model clusters with circular orbits at 6 kpc and 18 kpc that have orbital inclinations of $0^ \circ$, $22^\circ$, and $44^\circ$. We then study the combined effects of orbital eccentricity and inclination by simulating a cluster with an orbital eccentricity of 0.5, orbiting between $R_p$ of 6 kpc and $R_a$ of 18 kpc, with the same orbital inclinations. We selected these orbital parameters to allow us to compare a cluster with an eccentric orbit to clusters with circular orbits at $R_p$ and $R_a$ over a range of inclinations. The models with orbital inclinations of $0^\circ$ were first introduced in \citet{webb14}, hence we refer the reader to that study for additional details on the input parameters used in our models.

The clusters orbit within a Milky Way-like potential made up of a $1.5 \times 10^{10} M_\odot$ point mass bulge (Equation \ref{bulge}), a \citet{miyamoto75} disk (Equation \ref{disk} with $M_d = 5 \times 10^{10} M_\odot$, a= 4.5 kpc, and b = 0.5 kpc), and a logarithmic halo potential (Equation \ref{halo} \citep{xue08}). The halo is scaled such that the three potentials combine to give a circular velocity ($v_C$) of 220 km/s at a galactocentric distance of 8.5 kpc in the plane of the disk \citep{aarseth03}. Therefore $R_C$ in Equation \ref{halo} is 8.8 kpc. The initial radius in the plane of the disk ($R_{xy}$), initial height above disk ($Z_i$), $R_p$, eccentricity and orbital inclination of each cluster is given in Table \ref{table:modparam}. Note that model names are based on orbital eccentricity (e.g. e05), circular orbit distance or apogalactic distance (e.g. r18) and orbital inclination (e.g. i22).

\begin{equation} \label{bulge}
\Phi_{bulge}(R_{gc})=\frac{-GM_{b}}{R){gc}}
\end{equation}

\begin{equation} \label{disk}
\Phi_{disk}(R_{xy},z)=\frac{-GM_{d}}{\sqrt{R_{xy}^2+[a+\sqrt{b^2+z^2}]^2}}
\end{equation}

\begin{equation} \label{halo}
\Phi_{halo}(R_{gc})=\frac{1}{2}(v_C^2)LOG(R_{gc}^2.0+R_C^2.0)
\end{equation}

\begin{table}
  \caption{Model Input Parameters}
  \label{table:modparam}
  \begin{center}
    \begin{tabular}{lccccc}
      \hline\hline
      {Model Name} & {$R_{xy}$} & {$Z_i$} & {$R_p$} & {e} & {i} \\
      { } & {kpc} & {kpc} & {kpc} & { } & {degrees} \\
      \hline

e0r6i0 & 6 & 0 & 6 & 0 & 0 \\
e0r6i22 & 5.56 & 2.25 & 6 & 0 & 22 \\
e0r6i44 & 4.32 & 4.17 & 6 & 0 & 44 \\
e05r18i0 & 6 & 0 & 6 & 0.5 & 0 \\
e05r18i22 & 5.56 & 2.25 & 6 & 0.5 & 22 \\
e05r18i44 & 4.32 & 4.17 & 6 & 0.5 & 44 \\
e0r18i0 & 18 & 0 & 18 & 0 & 0 \\
e0r18i22 & 16.69 & 6.74 & 18 & 0 & 22 \\
e0r18i44 & 12.73 & 12.73 & 18 & 0 &44 \\

      \hline\hline
    \end{tabular}
  \end{center}
\end{table}

\begin{figure}
\centering
\includegraphics[width=\columnwidth]{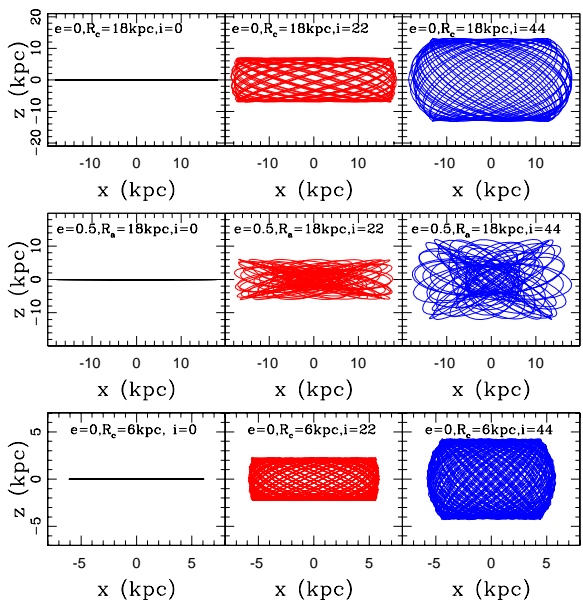}
\caption{Orbits of all model clusters. Clusters with circular orbits at 6 kpc are in the lower row, with orbital eccentricities of 0.5 and perigalactic distances of 6 kpc in the middle row, and circular orbits at 18 kpc in the top row. Orbital inclination changes from $0^\circ$ in the left column, to $22^\circ$ in the middle column, to $44^\circ$ in the right column.}
\label{fig:orbits}
\end{figure}

In order to better visualize the orbits of each model cluster, specifically how they evolve with time, we have plotted the x and z coordinates at each time step for each cluster in Figure \ref{fig:orbits}. The orbital eccentricity, circular orbit distance or apogalactic distance, and orbital inclination are marked in each panel.

\section{Influence of Orbital Inclination \label{sthree}}

To study the effects of orbital inclination on the scale sizes of clusters, we focus on the evolution of the mass, tidal radius, velocity dispersion, limiting radius and half-mass radius of all bound stars in each model cluster. A star is considered to be bound if the difference between its kinetic energy and the potential energy due to all other stars in the simulation is less than 0. 

\subsection{Mass}

\begin{figure}
\centering
\includegraphics[width=\columnwidth]{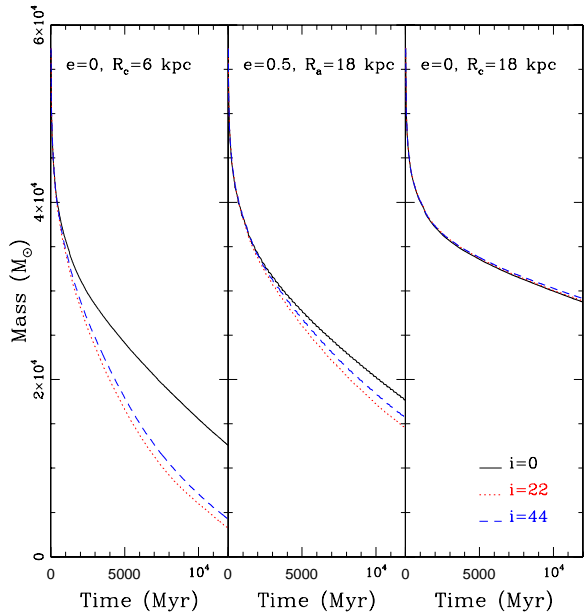}
\caption{The evolution of total cluster mass over time for clusters with a circular orbit at 6 kpc (left panel), an orbital eccentricity of 0.5 and an apogalactic distance of 18 kpc (center panel) and a circular orbit at 18 kpc (right panel). The black solid lines, red dotted lines, and blue dashed lines correspond to models with orbital inclinations of $0^{\circ}$, $22^{\circ}$, and $44^{\circ}$ respectively.}
\label{fig:m_t_b}
\end{figure}

The total bound mass of each cluster is plotted in Figure \ref{fig:m_t_b} as a function of time. For any cluster, mass loss is driven by stellar evolution and the tidal stripping of stars pushed beyond $r_t$. For clusters with circular orbits in the plane of the disk, the mean mass loss rate increases with decreasing $R_{gc}$ due to the increased strength of the tidal field. For clusters which experience non-static tidal fields (those with eccentric and/or inclined orbits) tidal heating and tidal shocks due to a sudden increase in the local gravitational potential (passage through a galactic disk or near $R_p$) are additional sources of mass loss. For example, clusters with circular orbits at 6 kpc that are inclined lose mass at a higher rate than the $i=0$ case, with e0r6i22 losing mass the fastest.

Model e0r6i22 ($22^{\circ}$ inclination) loses more mass than e0r6i44 ($44^{\circ}$ inclination) over 12 Gyr for two reasons. First, e0r6i22 passes through the disk more often, thus experiencing more frequent disk shocking. Secondly, the tidal field of the disk is proportional to $z^{-1}$ for a given $R_{xy}$ (see Equation \ref{disk}, so the cluster e0r6i22 spends the majority of its time in a stronger tidal field. e0r6i44 is far enough from the plane of the disk that when at its maximum height $z_{max}$ it experiences a weaker and nearly spherically symmetric tidal field such that tidal heating is less of a contributing factor. Model clusters with higher inclinations, like those performed by \citet{madrid14}, are also in agreement with our findings. Clusters with extremely high inclinations not only pass through the disk more frequently due to shorter orbital periods, but also pass through the disk perpendicular to the Galactic plane. Crossing the disk at such a high inclination increases the amount of energy imparted to cluster stars. Stronger and more frequent disk shocks experienced by high inclination clusters result in an accelerated mass loss rate compared to the models presented here.

For circular orbits at 18 kpc, there is very little difference between the mass profiles of the inclined and non-inclined cases. The strength of a \citet{miyamoto75} disk decreases as ${R_{xy}^{-1}}$ (Equation \ref{disk}). Hence for clusters orbiting at similarly large distances, the majority of the disk's mass is within their orbit, and the clusters evolve more as if they are in a spherically symmetric potential. However it is surprising that the clusters on inclined orbits are actually more massive at all times than the $i=0$ case since we expect these clusters to undergo some degree of disk shocking and tidal heating. This will be addressed in Section 4.2.

For clusters with eccentric orbits, periodic episodes of enhanced mass loss due to perigalactic passes are present in all three cases. In the case of a cluster with orbital eccentricity e in the plane of the disk (e05r180), the cluster takes (1+e) times longer to reach dissolution than a cluster with a circular orbit at $R_p$, or (1-e) times shorter than a cluster with a circular orbit at $R_a$. The dissolution time scaling is in agreement with \citet{baumgardt03}, who defines the dissolution time as the time it takes for the cluster to reach $100 M_\odot$. Given that the behaviour of collisional $N$-body simulations can become noisy at late times when only a small number of stars remain, we have checked the results of Figure \ref{fig:m_t_b} against an alternative definition of the dissolution time (when the cluster reaches $10 \%$ of its original mass) and find no noticeable change.

The amount of mass lost during a perigalactic pass decreases with increasing inclination because the tidal field is weaker at $R_p$ when the cluster is above or below the plane of the disk. However, the inclined and eccentric clusters still lose more mass than e05r18i0 because they undergo additional mass loss via disk shocking and increased tidal heating. While tidal heating has been shown to be a factor for clusters with eccentric orbits in the plane of the disk \citep{webb13}, it is even more effective for clusters with inclined orbits as the rate of change of the local potential is higher. The rate of change of the local potential is reflected in plots of $r_t$ versus time and the cluster's height above the disk as discussed in Sections 3.2 and 4.1 respectively. Disk shocking, while still an additional source of mass loss, is less effective than if the cluster had a circular orbit at $R_p$ because disk passages occur at larger galactocentric radii. The combined effects result in the (1-e) scaling factor from the perigalactic case remaining an accurate indicator of dissolution time (within $9\%$) for a given orbital inclination, while the $R_a$ case does not (greater than $30\%$).

\subsection{Tidal Radii}

\begin{figure}
\centering
\includegraphics[width=\columnwidth]{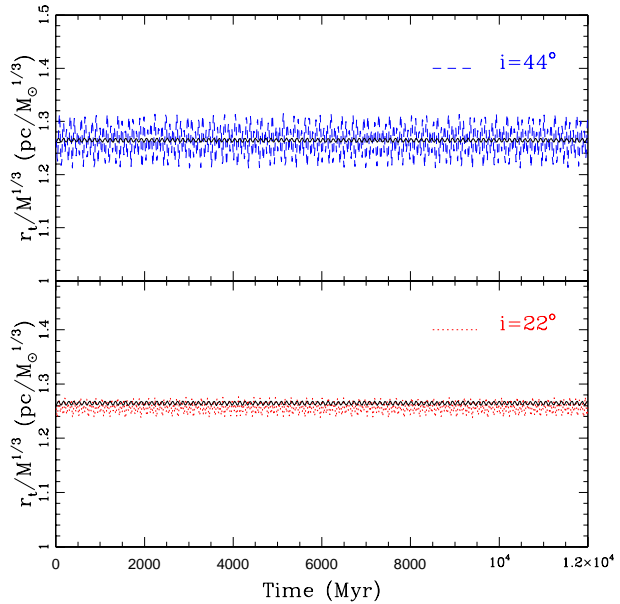}
\caption{The evolution of the mass normalized tidal radius over time for clusters with a circular orbit at 6 kpc. The black solid lines, red dotted lines, and blue dashed lines correspond to models with orbital inclinations of $0^{\circ}$, $22^{\circ}$, and $44^{\circ}$ respectively.}
  \label{fig:rt_t_b}
\end{figure}

Tidal shocks and tidal heating can be traced by the evolution of $r_t$ over the course of a cluster's orbit and lifetime. Any correlation between $r_t$ and orbital phase indicates that tidal heating is occurring, while a tidal shock occurs when $r_t$ suddenly goes from decreasing to increasing. To illustrate events of tidal shocking and heating, we plot the instantaneous $r_t$ of the models with circular orbits at 6 kpc in Figure \ref{fig:rt_t_b}. Model e0r6i22 is plotted in the lower panel (red) and e0r6i44 in the upper panel (blue). The non-inclined case, e0r6, has been plotted in black in both panels. The instantaneous $r_t$ has been calculated via Equation \ref{rt} given each cluster's mass and instantaneous location in the Galactic potential. To remove any dependence of $r_t$ on the mass loss rate and focus on effects due to cluster orbit, $r_t$ has been normalized by $M^{\frac{1}{3}}$ (See Equation \ref{rtHoerner}). Hence for clusters with circular orbits in the plane of the disk, their mass normalized $r_t$ never changes.

\begin{figure}
\centering
\includegraphics[width=\columnwidth]{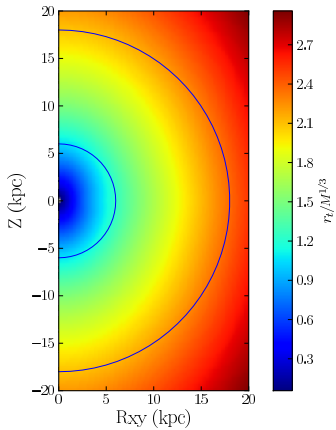}
\caption{The mass normalized instantaneous tidal radius at all points in the $R_{xy}-z$ plane. Solid lines mark galactocentric distances of 6 kpc and 18 kpc.}
  \label{fig:rtsurface}
\end{figure}

Figure \ref{fig:rt_t_b} indicates that the $r_t$ of clusters with inclined orbits fluctuates by $\pm 5 \%$ over the course of a single inclined orbit. The fluctuations in $r_t$ can be understood by plotting the mass normalized $r_t$ at all locations in the Galactic potential with $R_{xy} < 20$ kpc and $|z| < 20 $ kpc in Figure \ref{fig:rtsurface}. A cluster will have its largest $r_t$ when at its maximum height above the disk, with $r_t$ decreasing as the cluster approaches the disk. The process is then reversed as the cluster leaves the disk again on its way to its maximum distance below the disk.

For clusters orbiting at 18 kpc we see that the tidal field is essentially spherically symmetric. The tidal field imposed by the Galactic disk alone, and its gradient, become independent of z at approximately 15 kpc. Therefore a possible cut-off radius for the influence of orbital inclination may exist. Future studies on how this cut-off radius may depend on initial cluster conditions or the assumed structure of the Galactic disk are planned.

For a cluster on an inclined \textit{and} eccentric orbit, $r_t$ will also be growing and shrinking as it moves towards and away from $R_p$. With the distance above or below the disk at $R_p$ and $R_a$ changing from one orbit to the next, a cluster with such a complicated orbit cannot be considered to be in any form of equilibrium, but is instead in a constant state of flux.

\subsection{Velocity Dispersion}

\begin{figure}
\centering
\includegraphics[width=\columnwidth]{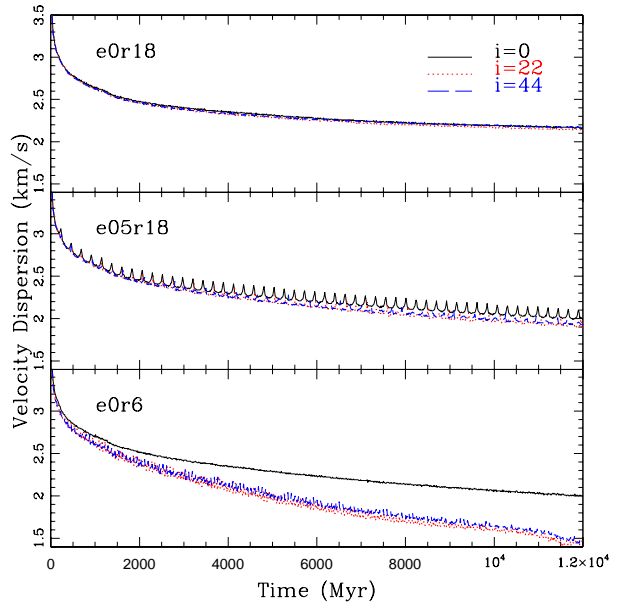}
\caption{The evolution of the velocity dispersion of all bound stars over time for clusters with a circular orbit at 6 kpc (bottom panel), an orbital eccentricity of 0.5 and an apogalactic distance of 18 kpc (center panel) and a circular orbit at 18 kpc (top panel). The black solid lines, red dotted lines, and blue dashed lines correspond to models with orbital inclinations of $0^{\circ}$, $22^{\circ}$, and $44^{\circ}$ respectively.}
  \label{fig:sig_t_b}
\end{figure}

The clearest demonstration of how these model clusters are affected by tidal shocks and tidal heating is in the evolution of the global three dimensional velocity dispersion $\sigma_V$ of all bound stars in Figure \ref{fig:sig_t_b}. The general trend in all cases is for $\sigma_V$ to decrease with time as mass segregated low-mass stars with higher velocities escape and the cluster loses mass. For a cluster on a circular orbit in the plane of the disk, the decrease in $\sigma_V$ is smooth. Periodic spikes in $\sigma_V$, that are only present in the inclined and eccentric clusters, are points where a sudden injection of energy (a tidal shock) has occurred.  A sudden increase in energy can cause a significant increase in stellar velocities \citep{webb14}.

For models on circular inclined orbits, each peak in $\sigma_V$ signifies a disk shock. The peak is followed by a sharp decrease in $\sigma_V$ as $r_t$ decreases and stars with lower binding energies (and high velocities) escape. $\sigma_V$ then slowly increases due to both the recapturing of temporarily unbound stars as $r_t$ begins to re-expand, and tidal heating as the cluster moves through a non-static tidal field. Hence the cluster expands as it moves towards $z_{max}$. The process then repeats itself when the clusters moves through the disk during the second half of its orbit. The strength of the shock decreases with $R_{gc}$ as the disk's contribution to the Galactic potential decreases.

The situation is slightly more complicated for models with eccentric \textit{and} inclined orbits like e05r18i22 and e05r18i44 . The cluster still crosses the disk twice per orbit, but since the orbit is non-circular the disk passages occur at different galactocentric radii. The disk shock that occurs nearest $R_p$ will be much stronger than the shock occurring near $R_a$. The weaker second shock results in the mass loss rate of inclined and eccentric clusters being only marginally higher than the non-inclined case. It is also important to note that when a cluster has a circular and inclined orbit, tidal heating and the recapturing of unbound stars is able to increase $\sigma_V$ between shocks. But when the cluster has an eccentric and inclined orbit, the weaker tidal field experienced as the cluster moves towards $R_a$ only injects enough energy to keep $\sigma_V$ constant between shocks. Therefore clusters on eccentric and inclined orbits do not expand in size as efficiently as clusters on inclined and circular orbits near $R_p$ while the cluster moves towards $z_{max}$.

\subsection{Limiting Radii}

\begin{figure}
\centering
\includegraphics[width=\columnwidth]{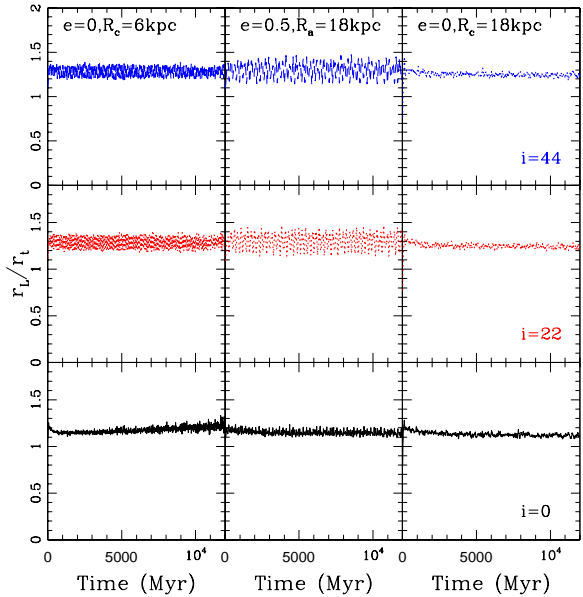}
\caption{The ratio of the limiting radius of each cluster to its tidal radius as a function of time for clusters with a circular orbit at 6 kpc (left panels), an orbital eccentricity of 0.5 and an apogalactic distance of 18 kpc (center panels) and a circular orbit at 18 kpc (right panels). The black (bottom panels), red (middle panels), and blue (top panels) lines correspond to models with orbital inclinations of $0^{\circ}$, $22^{\circ}$, and $44^{\circ}$ respectively.}
  \label{fig:rtrat}
\end{figure}

We next consider the effect of inclination on the \textit[limiting radius]. For the purposes of this study, the limiting radius is defined as the average cluster-centric distance of all bound stars located beyond the instantaneous $r_t$ \citep{webb13}. The interplay between a changing theoretical $r_t$ and the actual size of the cluster $r_L$ is illustrated in Figure \ref{fig:rtrat} where we plot the ratio of $\frac{r_L}{r_t}$ as a function of time. Based on our definition of $r_L$, the ratio will always be slightly larger than 1.0. 

How a cluster responds to its instantaneous $r_t$ is indicated by how much the ratio fluctuates around its mean value. For example, it has been shown that clusters which orbit in the plane of the disk fill their instantaneous $r_t$ at all times \citep{webb13}, which is why the ratio is nearly constant as a function of time for the non-inclined cases. The inclined cases, however all fluctuate around the mean $\frac{r_L}{r_t}$ value of the non-inclined case. The fluctuations are not the result of a non-static field, as e05r18 has a nearly constant ratio despite orbiting between 6 kpc and 18 kpc in the plane of the disk. The oscillations are due to inclined clusters being subject to increased tidal heating and additional shocking events per orbit compared to clusters in the plane of the disk. Before the cluster even has a chance to respond to its new local potential, which takes approximately one crossing time \citep{madrid14}, the local potential has already changed so quickly that the cluster never comes to equilibrium. 

During each orbit, when the cluster is moving away from the disk and towards $z_{max}$ it will be slightly underfilling as $r_t$ expands. As the cluster approaches $z_{max}$ it slows down and therefore has time to respond to its local potential and fill $r_t$. As the cluster moves away from $z_{max}$ and towards the plane of the disk, the cluster is slightly overfilling since $r_t$ is now decreasing. As the cluster passes through the plane of the disk and undergoes a disk shock, outer stars can become permanently or temporarily unbound, and the cluster becomes briefly tidally filling before $r_t$ begins to increase again.

\subsection{Half-mass Radius}

The inner structure of globular clusters, observationally traced by the effective radius $r_h$, is far more robust and less model dependent than $r_L$ \citep[e.g.][]{mclaughlin05, webb12, puzia14}. For $N$-body simulations, the half-mass radius $r_m$ is more commonly used to probe the inner regions of globular clusters \citep[e.g.][]{gieles10, madrid12, webb13}. The three-dimensional half mass radius is taken to be the radius enclosing half of the total bound mass, including both bound objects orbiting beyond $r_t$ and stellar remnants. The latter points resulting in $r_m$ being on average slightly larger than $r_h$.

\begin{figure}
\centering
\includegraphics[width=\columnwidth]{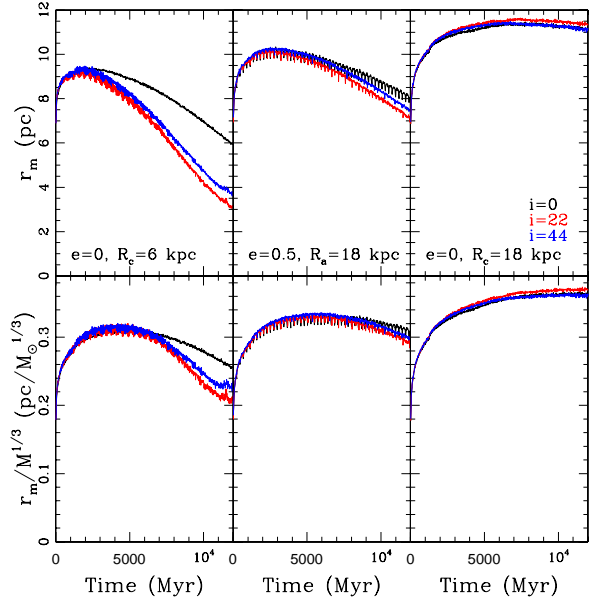}
\caption{The evolution of the half mass radius (top panels) and the half mass radius normalized by mass (bottom pannels) over time for clusters with a circular orbit at 6 kpc (left panels), an orbital eccentricity of 0.5 and an apogalactic distance of 18 kpc (center panels) and a circular orbit at 18 kpc (right panels). The black solid lines, red dotted lines, and blue dashed lines correspond to models with orbital inclinations of $0^{\circ}$, $22^{\circ}$, and $44^{\circ}$ respectively.}
  \label{fig:rm_t_b}
\end{figure}

 The half-mass radius and the mass normalized half-mass radius of each cluster as a function of time are plotted in Figure \ref{fig:rm_t_b}. It should be noted that our clusters all have final half-mass radii 2-3 times greater than most actual globular clusters. Future studies will explore the influence of inclined orbits in disk potentials on a wider range of initial $r_m$.

Figure \ref{fig:rm_t_b} suggests that the inner structure of a star cluster is less affected by changes in orbital inclination than $r_L$. If we first consider the models orbiting at 6 kpc, the inclined clusters are smaller because they lose mass at a faster rate than the non-inclined case. However if we normalize by mass, the mass normalized $r_m$ of all three cases are nearly identical for almost 7 Gyr. At 7 Gyr, the inclined clusters are approximately $1 \times 10^4 M_\odot$ in mass, and are in the process of dissolving. The eccentric clusters only differ in $r_m$ by 1 pc after 12 Gyr and the clusters orbiting at 18 kpc differ by less than 0.5 pc. After normalizing by cluster mass, the eccentric and 18 kpc clusters are nearly always identical in size.

\section{Discussion}

Our simulations indicate that the primary effect of an inclined orbit in a non-spherically symmetric potential is an increased mass loss rate due to tidal heating and shocking. To apply these findings to observations of star clusters, we need to know how tidal shocks and heating depend on cluster orbit and how they would influence the calculated size of a cluster.

\subsection{Tidal Heating and Shocks}

\begin{figure}
\centering
\includegraphics[width=\columnwidth]{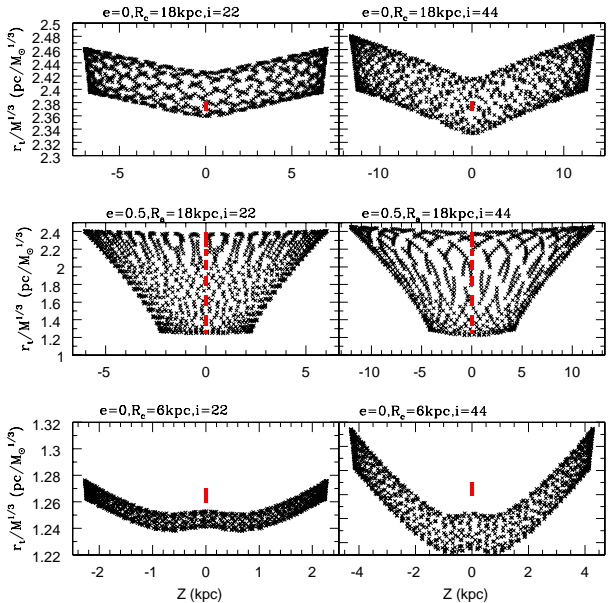}
\caption{Mass normalized tidal radius as a function of height above the disk z for all inclined model clusters (black crosses). Data points are equally spaced in time and cover 12 Gyrs of evolution, so the density of points reflects the amount of time the cluster spends at a given z. Red squares mark model clusters with orbits in the plane of the disk.}
  \label{fig:rt_z_norm}
\end{figure}

The increased mass loss rate experienced by clusters on inclined orbits is a direct result of them being subject to both tidal heating and tidal shocks, neither of which clusters on non-inclined circular orbits experience. We examine both of these effects further by plotting the mass normalized $r_t$ of each model cluster as a function of its height z over 12 Gyrs (Figure \ref{fig:rt_z_norm}). Since data points are equally spaced in time, the density of points reflects the proportion of its lifetime a cluster spends at a given z.

In Figure \ref{fig:rt_z_norm}, episodes of tidal heating are indicated by gradual changes in $r_t$. All model clusters experience some degree of tidal heating as $r_t$  decreases while the cluster moves inward from $z_{max}$. Tidal shocks are seen in Figure \ref{fig:rt_z_norm} when the mass normalized $r_t$ goes from decreasing to increasing. For an inclined orbit that is perfectly circular (e=0), a disk shock would be a singular event as $r_t$ reaches a minimum at $z=0$. However, since the orbits of our model clusters at 6 kpc and 18 kpc are not perfectly circular ($e \le 0.05$), the situation is slightly more complicated. At smaller $R_{xy}$ (6 kpc) the \textit{disk shock} appears to consist of two shocking events, just before and just after the cluster passes through the plane of the disk, unless the disk passage actually occurs at $R_p$. The increase in $r_t$ to a local maximum between the shocks lets the cluster temporarily expand freely. The dual shocks are what separates a disk shocking event from a more simple tidal shock, like a perigalactic pass, and make it more efficient at removing stars from the cluster. However, since the shape of the disk potential changes and strength of the disk decreases with $R_{xy}$, both shocks are not necessarily equal in magnitude, with one of the shocks sometimes being weaker and even negligible when orbits are near circular. At larger $R_{xy}$ (18 kpc), the decreased strength of the disk results in the disk shock being a singular event. Inclined \textit{and} truly eccentric clusters (middle row of Figure \ref{fig:rt_z_norm}) experience a strong dual tidal shock during its innermost disk passage, a weaker single shock during its outermost disk passage, and a third shock during each perigalactic pass.

Figure \ref{fig:rt_z_norm} indicates that inclined, eccentric clusters experience varying amounts of tidal heating from one orbit to another. While some orbits  keep the cluster at a high z until just before crossing the disk (minimizing tidal heating), other orbits gradually bring the cluster in from $z_{max}$ to z=0 (maximizing tidal heating). The complex orbits of clusters that are inclined and eccentric makes quantifying the effects of tidal heating or shocking difficult. However this study suggests that the evolution of a cluster with an eccentric and inclined orbit is more similar to a cluster with the same eccentricity and $i=0^{\circ}$ rather than a cluster with the same i orbiting at $R_p$ with e=0.

\subsection{The Effective Tidal Radius of an Inclined Orbit}

Because of the chaotic evolution of the instantaneous $r_t$ of a cluster with an inclined orbit, and because $r_L$ does not precisely trace $r_t$ as in a spherically symmetric potential, it is difficult to define what exactly the \textit{size} of a star cluster is. While an inclined cluster may appear to have an $r_L$ greater than its current $r_t$, this could simply be a function of its current orbital phase and not accurately indicate its current dynamical state. As we saw in Figure \ref{fig:rtrat}, an inclined cluster ranges between being tidally over-filling and under-filling, except at $z_{max}$ and $z=0$ when $r_L$ and $r_t$ are near equal.

\begin{figure}
\centering
\includegraphics[width=\columnwidth]{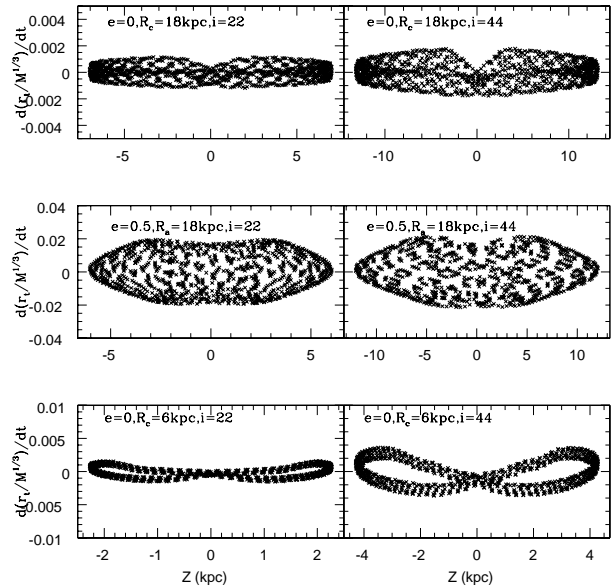}
\caption{The rate of change in the mass normalized tidal radius ($pc/M_\odot s^{-1}$ as a function of height above the disk z for all inclined model clusters. Data points are equally spaced in time and cover 12 Gyrs of evolution, so the density of points reflects the amount of time the cluster spends at a given z.}
  \label{fig:drt_z_norm}
\end{figure}

We wish to define an \textit{effective} $r_t$ for a cluster with an inclined orbit, in order to get a sense of its dynamical state and whether or not the cluster is tidally filling. When defining an \textit{effective} $r_t$, it should ideally represent a  stable state during the cluster's orbit at which the cluster spends the majority of its orbit. We consider the rate of change in the mass normalized tidal radius as a function of height above the disk in Figure \ref{fig:drt_z_norm}. The $r_t$ of a cluster near the plane of the disk fluctuates dramatically during the disk passage, and only represents a brief portion of the total orbit. Setting the \textit{effective} $r_t$ equal to the $r_t$ near $z=0$ would be equivalent to setting the $r_t$ of a cluster on an eccentric orbit equal to its $r_t$ at $R_p$, which we know to be incorrect \citep{webb13}. The clear choice is to let the \textit{effective} $r_t$ of a cluster with an inclined orbit be equal to its instantaneous $r_t$ at $z_{max}$. Not only is the rate of change in $r_t$ is at its minimum when the cluster is both approaching and leaving $z_{max}$, but inclined clusters also spend the majority of their lifetime near $z_{max}$. Therefore the time averaged $r_t$ of each model also corresponds to $r_t$ at $z_{max}$. 

\section{Summary}

In this paper, we simulated the evolution of star clusters orbiting in a Milky Way-like potential with a range of orbital inclinations in order to study the effects of orbital inclination on their dynamical and structural evolution. The main factors which separate star clusters on inclined orbits are tidal heating and tidal shocking. While clusters on eccentric orbits in a spherically symmetric potential do experience tidal heating and tidal shocking at perigalacticon, both inclination and eccentricity are more dominant when the orbit is inclined and a galactic disk is present. The strength of both tidal heating and shocking due to an inclined orbit however weakens with $R_{gc}$. By 18 kpc, the Galactic potential is nearly spherically symmetric and orbit inclination is nearly negligible. When performing $N$-body simulations of remote halo clusters, such as Pal 4 and Pal 14 \citep{zonoozi11, zonoozi14}, unless their orbits are highly eccentric and bring them deep into the inner regions of the Milky Way it can be safely assumed that they orbit in a spherically symmetric potential.

We have simulated model clusters with identical initial conditions with both circular and eccentric orbits over a range of orbital inclinations to determine the main effects of tidal heating and tidal shocking. Our main conclusions are as follows:

\begin{itemize}

\item  For clusters with small $R_{gc}$, inclined clusters experience an enhanced mass loss rate due to increased tidal heating and two tidal shocking events during a disk passage. Clusters with small orbital inclinations are more strongly affected since they spend a longer time in the stronger disk potential. 
\item At higher $R_{gc}$, the strength of the galactic disk is weaker, minimizing the effects of tidal heating and disk shocking. Furthermore, $r_t$ at $z_{max}$ is larger than in the plane of the disk, so inclined clusters will actually lose mass at a lower rate than non-inclined clusters.
\item Disk shocking causes a temporary increase in $\sigma_V$, followed by a sharp drop as stars that have been energized to higher velocities escape the cluster.
\item Between shocking events, $\sigma_V$ can remain constant or even increase due to tidal heating.
\item The local potential around a cluster with an inclined orbit is in a constant state of flux, so an inclined cluster is not able respond to its instantaneous $r_t$ except at $z_{max}$. The $r_L$ of the cluster instead fluctuates around $r_t$ at $z_{max}$, ranging between being tidally under-filling and over-filling as it travels away from or towards the disk respectively.
\item Tidal heating and shocking have a negligible effect on the inner region of the cluster ($r<r_m$).
\item The tidal radius of a cluster on an inclined (or inclined and eccentric) orbit is best approximated by assuming it has a circular orbit at its maximum height above the disk: $r_t(R_{xy},z,e,i) = r_t(R_{xy},z_{max})_{z_{max}}$

\end{itemize}

The final point that $r_m$ is unaffected by orbital inclination is helpful when studying globular clusters in other galaxies. More specifically in disk galaxies or elliptical galaxies that are triaxial, the commonly observed effected radius is independent of the orientation of the clusters orbit in the galactic potential, which would be difficult to determine. Therefore the effective radius is solely dependent on the cluster's three dimensional position and orbital eccentricity.

The combined effects of orbital inclination and eccentricity on a cluster are complex. The cluster experiences a strong disk shock when it crosses the disk near $R_p$, a weak disk shock when crossing near $R_a$, and a tidal shock during its perigalactic pass. Furthermore, the cluster does not cross the disk at the same $R_{gc}$ or reach $R_p$ at the same $z$ from one orbit to the next. The cluster is also constantly subjected to tidal heating since both the $R_{gc}$ and $z$ coordinate of the cluster change with time. Ultimately, predicting the evolution of a cluster with an inclined and eccentric orbit is difficult, although the effects of orbital inclination clearly decrease with increasing orbital eccentricity since high-e clusters spend the majority of their lifetime at large galactocentric radii. We do find that the dissolution time of such a cluster can be approximated to be (1+e) times longer than the dissolution time of a cluster with a circular orbit at $R_p$ and the same orbital inclination, in agreement with the work of \citet{baumgardt03} for clusters with non-inclined obits.

Exploring a large parameter space in both orbital inclination and eccentricity, and their subsequent effects on clusters, is necessary as the orbits of Galactic globular clusters are neither circular nor in the plane of the disk. As previously mentioned, we also wish to explore how the initial $r_m$ of a cluster changes its dynamical evolution in a non-spherically symmetric potential. The ultimate goal is to be able to predict the size of any cluster no matter its position or orbit in an arbitrary tidal field. Any clusters whose theoretical and observational sizes do not match may indicate recently captured clusters that have not spent long in their current tidal field. When theory and observations do match, we will be able to predict other dynamical properties of a cluster, including its stellar mass function, and rewind the cluster's dynamical clock to determine its initial mass and initial size. 

\section{Acknowledgements}

JW, WEH and AS acknowledge financial support through research grants and scholarships from the Natural Sciences and Engineering Research Council of Canada. This work was made possible by the facilities of the Shared Hierarchical 
Academic Research Computing Network (SHARCNET:www.sharcnet.ca) and Compute/Calcul Canada.


\bsp

\label{lastpage}

\end{document}